\documentclass[astra]{copernicus}

\begin{document}

\title{Antiproton modulation in the Heliosphere and AMS-02 antiproton over proton ratio prediction\\
\small{(accepted for publication)}}
\author[1]{P. Bobik}
\author[2,4]{M.J. Boschini}
\author[2]{C. Consolandi}
\author[2,5]{S. Della Torre}
\author[2,3]{M. Gervasi}
\author[2]{D. Grandi}
\author[1]{K. Kudela}
\author[2,3]{S. Pensotti}
\author[2]{P.G. Rancoita}

\affil[1]{Institute of Experimental Physics, Kosice (Slovak Republic)}
\affil[2]{Istituto Nazionale di Fisica Nucleare, INFN Milano-Bicocca, Milano (Italy)}
\affil[3]{Department of Physics, University of Milano Bicocca, Milano (Italy)}
\affil[4]{CILEA, Segrate (MI) (Italy) }
\affil[5]{Department of Physics and Maths, University of Insubria, Como (Italy)}


\runningtitle{Antiproton modulation in the Heliosphere and AMS-02 antiproton over proton ratio prediction}

\runningauthor{Davide Grandi}

\correspondence{Davide Grandi\\ (Davide.Grandi@mib.infn.it)}

\received{}
\pubdiscuss{} 
\revised{}
\accepted{}
\published{}


\firstpage{1}

\maketitle

\begin{abstract}
\noindent We implemented a quasi time-dependent
2D stochastic model of solar modulation describing the transport of cosmic rays
(CR) in the heliosphere.
Our code can modulate the Local Interstellar Spectrum (LIS) of a generic charged particle (light cosmic ions and electrons),
calculating the spectrum at 1AU.
Several measurements of CR antiparticles have been performed. Here we focused our attention on the
CR
antiproton component and the antiproton over proton ratio.
We show that our
model, using the same heliospheric parameters for both particles, fit the observed
$\frac{\bar{p}}{p}$ ratio.
We show a good agreement with BESS-97 and
PAMELA data and make a prediction for the AMS-02 experiment.
\end{abstract}


\introduction

Galactic cosmic rays (GCRs) are nuclei, with a small component of leptons,
mainly produced by supernova remnants [\cite{blasi2010}], confined by the galactic magnetic field
to form a isotropic flux inside the galaxy.
Before reaching the Earth orbit they enter the heliosphere, the region where the interplanetary magnetic
field is carried out by the solar wind (SW).
In this enviroment they undergo
diffusion, convection, magnetic drift and adiabatic
energy loss, resulting in a reduction of particles flux at
low energy ($\leq$1-10 GeV) depending on solar activity
and polarity.
This effect is known  as \textit{solar modulation}.
We have developed a
2D (radius and helio-colatitude) model of GCR
propagation [\cite{six}] in the heliosphere, by using stochastic differential equations (SDEs).
The model depends on
measured values of the SW velocity on the ecliptic plane ($V_0$),
tilt angle ($\alpha$) of the neutral sheet and estimated values
of the diffusion parameter ($k_{0}$):
details on parameters are discussed in section 2 and 3.
This model includes drift transport due to magnetic
field curvature and gradients, as well
the presence of a tilted neutral sheet describing properly periods of low and medium solar activity.
Modulated fluxes depend on solar activity but also
on particle charge and solar magnetic polarity
[\cite{art_midrift}].

\section{Stochastic 2D Monte Carlo code}

The GCR transport in the Heliosphere is described by a
Fokker-Planck equation, the so-called
Parker equation [\cite{parker1965}]:
\begin{eqnarray}
\frac{\partial U}{\partial t} &= & \frac{\partial}{\partial
x_i}\left( K^S_{ij}\frac{\partial U}{\partial x_j} \right)
-\frac{\partial}{\partial x_i}(V_{sw_i}U)
\nonumber\\ [5pt]
&& +\frac{1}{3}\frac{\partial V_{sw_i}}{\partial
x_i}\frac{\partial}{\partial T}(\varrho T U)
-\frac{\partial}{\partial x_i}(v_{D_i} U)
\label{eq_parker}
\end{eqnarray}
\noindent where $U$ is the cosmic ray number density per unit
interval of particle kinetic energy, $t$ is the time, $T$ is the
kinetic energy (per nucleon), $V_{sw_i}$ the SW velocity
along the axis $x_i$, $ v_{D_i} $ is the drift
velocity
related to the antisymmetric part of diffusion tensor
[\cite{jokipii77b}], $K^S_{ij}$ is the symmetric part of the
diffusion tensor and
$\varrho$=$(T+2T_{0})/(T+T_{0})$ [\cite{gleeson1967}],
where $T_{0}$ is particle's rest energy.
This
partial differential equation is equivalent [\cite{ito}] to a set of ordinary
SDEs that can be integrated
with Monte Carlo (MC) techniques. The integration time step ($\Delta
t$), is taken to be proportional to $r^{2}$ ($r$
is the distance from the Sun) avoiding oversampling in the outer heliopshere
and therefore saving CPU
time [\cite{Alanko_usoskin2007}]. We considered the 2D (radius and
colatitude) approximation of Eq.~\ref{eq_parker}, and from this we calculate the
equivalent set of SDEs:
\begin{eqnarray}
\Delta r &= &  \frac{1}{r^2}\frac{\partial (r^2 K_{rr})}{\partial
r}\Delta t + \left( V_{sw}+ v_{D_r}  +
v_{D_{NS}} \right)\Delta t \nonumber\\[5pt]
&& +R_g\sqrt{2K_{rr}\Delta t} \label{deltax}\\
\Delta \mu &= &   \frac{1}{r^2}\frac{\partial [(1-\mu^2)K_{\theta \theta}]}{\partial \mu}  \Delta t - \left(
\frac{\sqrt{1-\mu^2}}{r} v_{D_\theta}  \right)\Delta t
\nonumber\\ [5pt]
&& + R_g\sqrt{ \frac{2K_{\theta \theta} (1-\mu^2)}{r^2} \Delta t} \\
\Delta T & = & -\left( \frac{2}{3}   \frac{\varrho V_{sw} T}{r}\right) \Delta t \label{SDEeqs_comp}
\end{eqnarray}
\noindent where $\mu$=$\cos\theta$, with $\theta$ colatitude, and
$R_g$ is a gaussian distributed random number with unitary
variance. Here the drift velocity is split in regular drift
(radial drift $v_{D_r}$, latitudinal drift $v_{D_\theta}$) and neutral sheet drift ($v_{D_{NS}}$) as described
by \cite{12}.
The radial diffusion coefficient is $K_{rr}$=$K_{||}
cos^{2}\psi$+$K_{\perp r} sin^{2}\psi$  [\cite{7}], where $\psi$ is the angle
between radial versor and direction of the solar magnetic field described below. The
latitudinal coefficient is $K_{\theta\theta}$=$K_{\perp\theta}$ [e.g., see~\cite{9}].
We note as the perpendicular diffusion coefficient has two components,
one in the radial direction ($K_{\perp r}$) and one in the polar direction
($K_{\perp \theta}$) as shown in \cite{potgieter2000}.
 We define $(K_{\perp})_{0_i}$ as the ratio
between perpendicular and parallel diffusion coefficients, therefore
$K_{\perp i}$=$(K_{\perp})_{0_i} K_{||}$.
We fixed this value:
$(K_{\perp})_{0_r}$=0.05 while
$(K_{\perp})_{0_\theta}=f(\theta)(K_{\perp})_{0_r} $ 
(where $f(\theta)= 10$ close to the poles and
$f(\theta) = 1$ in the equatorial region)[\cite{potgieter2000}],
to reproduce the correct magnitude
and rigidity dependence of the latitudinal cosmic ray proton and electron gradients [cf.
\cite{potgieter97, burger2000}].
The parallel diffusion coefficient  is $K_{||}$=$k_{0}\beta
K_{P}(P)(B_{\oplus}/3B)$ [\cite{9}];
here $k_{0} \simeq 0.05-0.3 \times
10^{-3}$ AU$^{2}$GV$^{-1}$s$^{-1}$, is a diffusion parameter depending on the solar actvitiy (see section~\ref{par}),
   $\beta$ is the particle velocity, $P$ is
the CR particle's rigidity, $K_{P}$=$P$, $B_{\oplus}$ is the
value of heliospheric magnetic field at the Earth orbit, and $B$ is
the magnitude of the Heliospheric Magnetic Field (HMF) [\cite{12}]:
\begin{eqnarray}
\mathbf B = \frac{A}{r^2}(\mathbf e_r - \Gamma \mathbf e_\phi)[1-2H(\theta - \theta')]
\label{Bpark}
\end{eqnarray}
where $A$ is a coefficient that determines the
field polarity and allows $|\mathbf  B|$ to be equal to $ B_{\oplus}$,~i.e.,
the value of IMF at the Earth orbit; $\theta'$ is the polar angle determining the position of
the heliospheric current sheet (HCS)~[\cite{Jokipii1981}];
 $H$ is the \textit{Heaviside} function,
thus $[1-2\,H(\theta-\theta')]$ for the change of sign between the two regions - above and below the HCS - of the heliosphere;
 finally $\Gamma= \tan\psi\simeq\frac{\omega \,r \sin\theta}{V_{ sw}} $,
with $\psi$ the spiral angle.
The Parker
 field has been {\mbox modified} introducing a small latitudinal component
 $B_{\theta}$=$\frac{A}{r^2}(r/r_{0})\delta(\theta) $ with $\delta(\theta)=8.7\times 10^{-5}/ \sin \theta$,
thus allowing one to obtain $\nabla \cdot \mathbf B =0$ and a field
magnitude according to \cite{jk89}:
\begin{eqnarray}
 B = \frac{A}{r^2}\sqrt{1+\Gamma^{2}+\left(\frac{r}{r_{0}}\right)^{2}\delta^{2}}
\end{eqnarray}
\noindent that increases  the magnitude of the HMF in
the polar regions without a modification of the field topology.
This component produces a lower magnetic
drift velocity in this region,
with the effect of a lower CR penetration along polar field lines in
the inner part of the heliosphere [\cite{jk89}].
We use a SW broad smoothed profile according to Ulysses data
for periods of low solar activity [\cite{McComas2000}],
described by the
relation $V_{sw}(\theta) = V_{max}$ if $\theta\leq30^{\circ}$ or $\theta\geq150^{\circ}$ and
$V_{sw}(\theta) = V_{0}\cdot(1+|\cos \theta|)$ if $30^{\circ}<\theta<150^{\circ}$
where
$V_{0}$ is approximately 400 km/s and $V_{max}$ is 760 km/s.
Drift effects are included through analytical effective drift
velocities:
in the Parker spiral field we evaluated
drift due to gradient, curvature and neutral
sheet 
that modify the integration path inside the
heliosphere.
We adopted the approach of Potgieter and Moraal [\cite{PM1985} ], because it is
able to reproduce the effects of drift in both quiet and active solar
periods [a discussion on other models can be found in \cite{art_ste}].
In this model the drift coefficient is modified with a transition function that simulates
the effect of a wavy neutral sheet. The sharpness of this function is related to $\alpha$ angle,
expanding or shrinking the region of influence of neutral sheet drift.
As LIS, both for protons and antiprotons, we use the ones used in \cite{casaus}
and obtained from Galprop\footnote{\url{http://galprop.stanford.edu/webrun/}}.
\section{Parameters and Data Sets}\label{par}
\begin{figure}[hbt!]
\vspace*{2mm}
\begin{center}
\includegraphics[width=8.3cm]{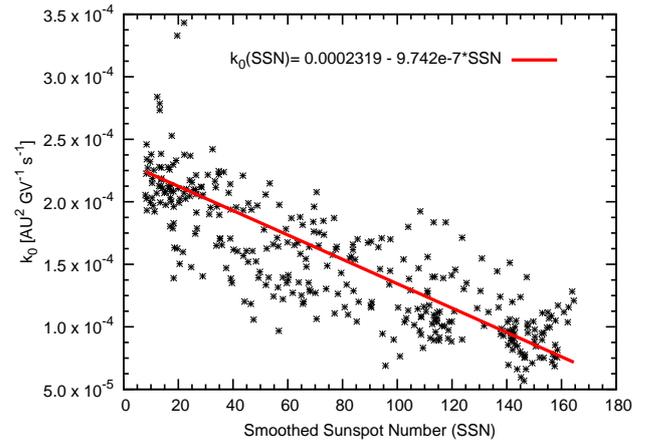}
\end{center}
\caption{$k_{0}$ values versus monthly SSN. The linear fit is also shown. Reported values cover the time period
1951-2004.}
\label{fig_sim1b}
\end{figure}
\noindent Values of the tilt
angle $\alpha$   are computed using two different models,
described in \cite{wso}, fitting separately periods of increasing solar activity and periods of decreasing solar activity
[\cite{ferreira2004}].
The three drift components do not
depend on external parameters, except the solar polarity, so A$>$0
for positive periods and A$<$0 for negative periods [\cite{six}].
We selected CR $p$ and $\bar{p}$ data from
several experiments in order to
compare and tune model results.
We modulated separately $p$ and $\bar{p}$ LIS spectra and then we computed the ratio.
In this paper we show experimental data taken during periods of low solar activity:
the comparison with BESS-97 (A$>$0 July 1997, see \cite{bess}) and PAMELA (A$<$0 from 2007 to 2008, see \cite{pamela}).
%
$V_0$ and $B_\oplus$ values for these periods were obtained from NSSDC OMNIWeb system\footnote{ \url{http://omniweb.gsfc.nasa.gov/form/dx1.html}}
by 27 daily averages, while tilt angle values from the Wilcox
Solar Laboratory [\cite{wso}].
We estimated the values of $k_{0}$, needed to evaluate the CR
modulation in different conditions, from the modulation parameter 
 reported in \cite{ilya05}.
We searched a relation between the estimated $k_{0}$ values
and the monthly Smoothed Sunspot Numbers (SSN).
We found that there is a nearly linear relation between $k_{0}$ and SSN\footnote{\url{http://www.sidc.oma.be/sunspot-data/}} values
(see Fig. \ref{fig_sim1b}),
with a Gaussian distribution of the best fit with a RMS of 19\%. This is a first crude estimation,
we will perform a more complex anaysis, e.g. fitting separately different solar phases, in order
to avoid systematics in the relation and to reduce the RMS.
 In this way we can use the estimated SSN values to obtain the
diffusion coefficient $k_{0}$. Following this approach we introduced in our code a gaussian
random variation of $k_{0}$ with a RMS of 19\%.
Results of the simulation with and without the gaussian variation
are consistent inside the indetermination of the code (around 5\%).
Our code simulates a diffusive propagation of a CR entering the heliosphere
from its outer limit, that we located
  at 100 AU (note that in \cite{pause} the Termination Shock is located at 94 AU) , and reaching
the Earth at 1 AU; the effects of heliosheath and termination shock are not taken into account in the present model.
We evaluated the time $t_{sw}$ needed by the SW to expand from
the outer corona up to 100 AU, with a minimum speed
of $\sim$400 km/s it takes nearly 14 months, while the time interval
$\tau_{ev}$ of the stochastic evolution of a quasi particle inside
the heliosphere from 100 AU down to 1 AU is between
1 month (at 200MeV) and few days (at 10 GeV). This scenario, where
$\tau_{ev} < t_{sw}$ and $t_{sw} >>$ 1 month, indicates that we
should use different parameters (monthly averages) to describe the
conditions of heliosphere in the modulation process. In fact at 100
AU, where particles are injected, the conditions of the solar
activity are similar to those present at the Earth $~$14
months before. Therefore, we  can divide the heliosphere in 14 regions as a function of
the radius. For each region we evaluated
$k_{0}$, $\alpha$
and $V_{sw}$, in relation to the time spent by the solar wind to reach this region.
We indicate the present treatment accounting for the time evolution of
the solar parameters
as a \textit{dynamic} approach of the heliosphere.

\section{Results}\label{sec:res}
\begin{figure}[hbt!]
\begin{center}
\includegraphics[width=8.3 cm]{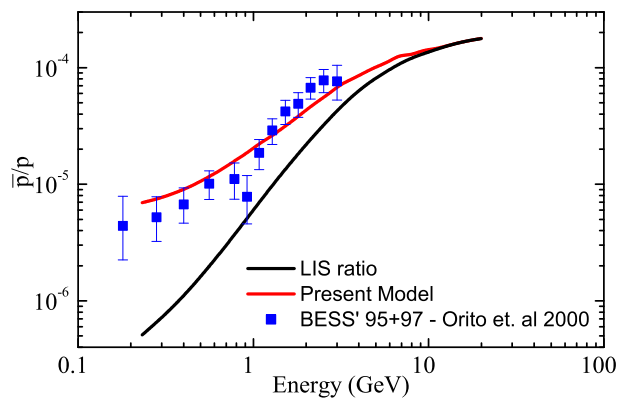}
\end{center}
\caption{Comparison of simulated $\frac{\bar{p}}{p}$ ratio at 1 AU and
experimental data: BESS (1997).}
\label{fig_sim1}
\end{figure}

\begin{figure}[hbt!]
\begin{center}
\includegraphics[width=8.3cm]{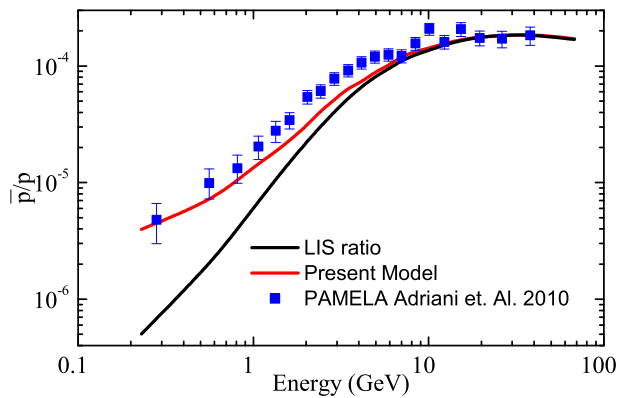}
\end{center}
\caption{Comparison of simulated $\frac{\bar{p}}{p}$ ratio at 1 AU and
experimental data: PAMELA (2007-2008).} \label{fig_sim2}
\end{figure}

Results obtained with our propagation code are shown in \mbox{Figs.
\ref{fig_sim1} and \ref{fig_sim2}}. Simulated fluxes obtained using parameters dependent on the
heliospheric region agree with measured data
within the experimental error bars.
 This happens both in periods with
A$>$0 (BESS-97), and in periods
with A$<$0 (PAMELA). This means that current
treatment of the Heliosphere improves the
understanding of the complex processes occurring inside the Solar Cavity.
\begin{figure}[hbt!]
\begin{center}
\includegraphics[width=8.3cm]{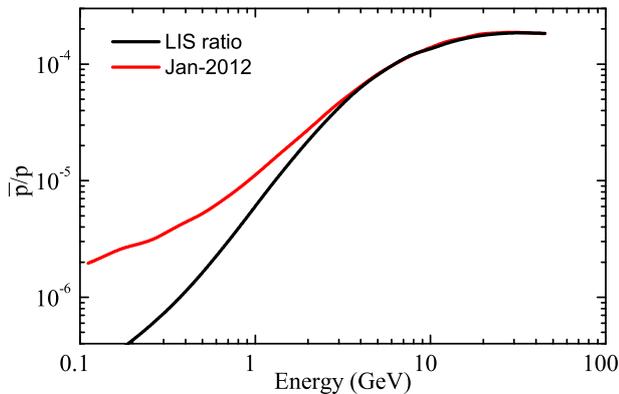}
\end{center}
\caption{Prediction of modulated $\frac{\bar{p}}{p}$ ratio at 1 AU for AMS-02.}
\label{fig_sim3}
\end{figure}
\noindent Our code can be also used to predict CR fluxes for future
measurements. The assumption is that diffusion
coefficient, tilt angle and solar wind speed show a near-regular
and almost periodic trend. The periodicity is two consecutive
11-years solar cycles. We selected periods with a similar
solar activity conditions and same solar field polarity of the time of interest:
 therefore approximately 22 years before.
We used the values measured in that
periods as an estimation of the conditions of the heliosphere.
Simulations have been carried out in prevision of the AMS-02
mission 
that will be installed on the ISS
in 2011: we choose January 2012.
For this period we show in Fig. \ref{fig_sim3} the predictions of GCR
modulation for the $\bar{p}/p$ ratio.
In order to reduce the uncertainty it is important
to compare our model with the AMS-02 data because of the huge
statistics and the long time covered. 

\section{Conclusions}\label{sec:conc}

We developed a 2D stochastic MC code for particles
propagation across the heliosphere.
We compared the ratios of
$\bar{p}/p$ fluxes  measured by BESS and PAMELA with those obtained
from the present MC code.
In the present calculations we used - for the parameters
$k_{0}$, $\alpha$ and $V_{sw}$ -
values corresponding to the periods of data taking. 
This
description
of the heliosphere and
the forward approach seem to
properly  account for the
propagation of GCR in the solar cavity.
Recent measurements [\cite{pamela}]
have pointed out the needs to reach a high level of accuracy in
the modulation of the fluxes, in relation to the charge sign of
the particles and the solar field polarity. 
 This
aspect will be even more crucial in the next generation of
experiments like AMS-02.












\end{document}